\documentclass[11pt,twoside]{article}
\usepackage{graphicx,epsfig,natbib,epstopdf}
\usepackage{CS18}
\begin{document}
%
%
%
\title{SOLIS: reconciling disk-integrated and disk-resolved spectra from the Sun}
\author{Alexei Pevtsov$^{1}$, Luca Bertello$^{2}$, Brian Harker$^{2}$, Mark Giampapa$^{2}$, Andrew Marble$^{2}$}
\affil{$^1$National Solar Observatory, Sunspot, New Mexico, USA 88349-0062}
\affil{$^2$National Solar Observatory, Tucson, Arizona, USA 85726-6732}
\begin{abstract}

Unlike other stars, the surface of the Sun can be spatially resolved to a high degree of detail. But the Sun can also be observed as if it was a distant star. The availability of solar disk-resolved and disk-integrated spectra offers an opportunity to devise methods to derive information about the spatial distribution of solar features from Sun-as-a-star measurements. Here, we  present an update on work done at the National Solar Observatory to reconcile disk-integrated and disk-resolved solar spectra from the Synoptic Optical Long-term Investigation of the Sun (SOLIS) station. The results of this work will lead to a new approach to infer the  information about the spatial distribution of features on other stars, from the overall filling factor of active regions to, possibly, the latitude/longitude distribution of features.
\end{abstract}
\section{Introduction}

Since the Sun can be observed both as if it was a distant a star and with sufficient spatial resolution allowing to map the solar surface to a high degree of detail, a direct comparison of two types of observations can be useful for interpreting stellar spectra. In particular, one can envision developing an approach to derive the fraction of the stellar surface occupied by a feature of a specific type (e.g., starspots, plages, network elements etc). \citet{Skumanich.etal1984} developed a three-components model representing the contributions from the interior of supergranular cells, network boundaries, and plage to match the spectral line profile of Ca {\rm II} K 393.37 nm observed at the minimum of the solar activity cycle. The model, however, was not successful in representing the disk-integrated line profiles during the rising phase of the solar cycle.
Based on the model spectra derived from 3D MHD modelling of solar convection, \citet{Uitenbroek.Criscuoli2011} concluded that simple one-dimensional models may not represent the average properties of stellar atmospheres even if they reproduce the mean observed spectra. They identified several possible reasons for this failure: nonlinearities (with temperature and/or density) in the Planck function and in molecular formation, as well as the anisotropy of convective motions. The latter strongly affects the center-to-limb variation of line-core intensities.
Despite these limitations, comparision of disk-integrated and disk-resolved spectra may  still be useful in answering simpler questions, such as the fraction of the stellar surface occupied by features of a specific class, which is the subject of the current study.

\begin{figure}[t!]
\centering
\includegraphics[width=0.9\textwidth]{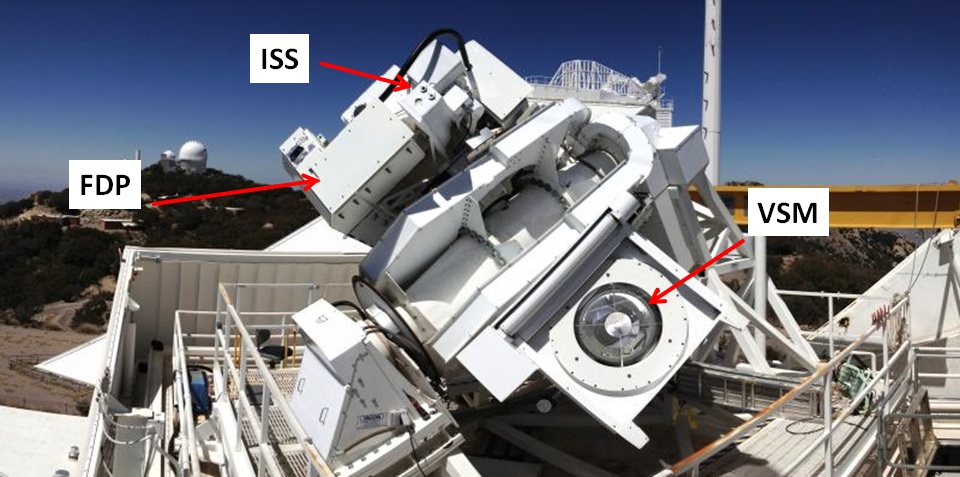}
\caption{SOLIS facility on Kitt Peak. Arrows mark the three main instruments: Vector Spectromagnetograph (VSM), Full Disk Patrol (FDP) and Integrated Sunlight Spectrometer (ISS).}
\label{fig:solis}
\end{figure}

\begin{figure}[h!]
\centering
\includegraphics[width=0.9\textwidth]{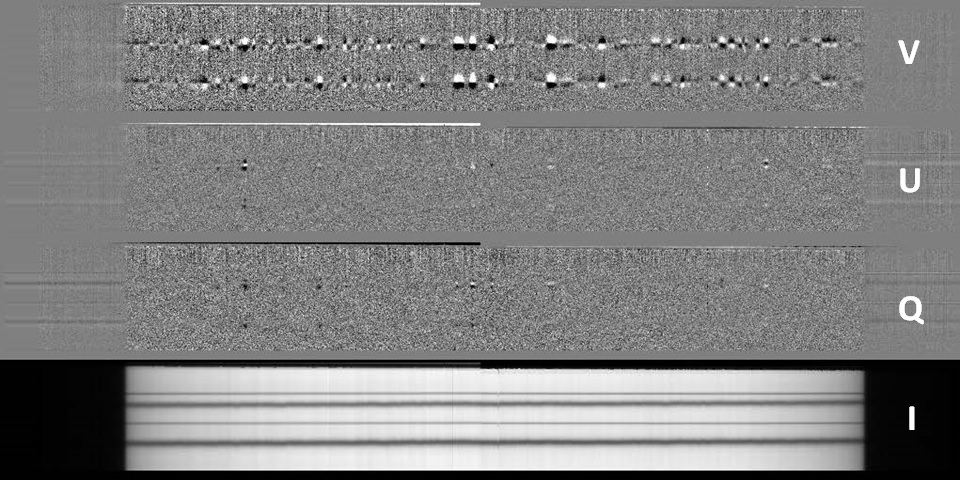}
\caption{Example of a single Fe {\rm I} 630.15-630.25 nm spectral scaneline taken with VSM. Dispersion is in the vertical direction, and spatial direction is horizontal (solar East limb is to the left and West limb is to the right). Spectra corresponding to four Stokes components are shown: I -- non-polarized light, V is circularly polarized light, and Q and U correspond to linearly polarized light. The two broader horizontal  lines in Stokes I correspond to the photospheric iron spectral lines, and the two narrower lines are the telluric lines of molecular oxygen. The black and white pattern in Stokes V delineates the presence of longitudinal magnetic field at each pixel along the spectrograph slit.}
\label{fig:spectra}
\end{figure}

\section{Observations}

We employ observations from the Synoptic Optical Long-term Investigations of the Sun (SOLIS) facility \citep{Keller.Harvey.Giampapa2003,Balasubramaniam.Pevtsov2011}, which has been in operation since 2003 at the NSO facilities on Kitt Peak, Arizona. SOLIS (Figure \ref{fig:solis}) is composed of a single equatorial mount carrying three instruments: the 50 cm Vector Spectromagnetograph (VSM), the 14 cm Full-Disk Patrol (FDP, a full disk tunable filter-based imager), and  the 8 mm Integrated Sunlight Spectrometer (ISS).

\begin{figure}[h]
\centering
\includegraphics[width=0.6\textwidth]{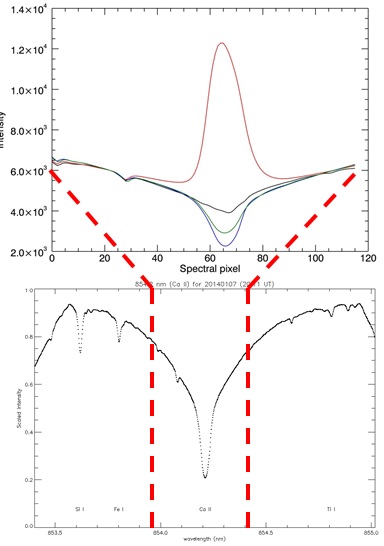}
\caption{Example of Ca {\rm II} 854.21 nm spectral line profiles from ISS (Sun-as-a-star, bottom) and VSM (top). Thick dashed lines mark the approximate spectral range observed by VSM. Examples of VSM spectra corresponding to different features are shown by different colors; from top to bottom: plage (red), large sunspot (black), quiet Sun at solar disk center (green), and quiet Sun at about 20 degrees of heliocentric distance (blue). For comparision, we scaled all profiles by mean intensity near the telluric line  (pixels 20 -- 25).}
\label{fig:spectra2}
\end{figure}

VSM uses a long (curved) slit to scan the solar image in declination, thus constructing a map of the solar surface 2048 $\times$ 2048 pixels in size, with 1 pixel corresponding to about 1.14 $\times$ 1.14 arcseconds.  For each position along the spectrograph slit, VSM records full line profiles in Fe {\rm I} 630.15-630.25 nm (photosphere, Stokes I, Q, U, and V), Ca {\rm II} 854.21 nm (chromosphere, Stokes I and V), and He {\rm I} 1083.0 nm (upper chromosphere, Stokes I only). It takes about 20/40 minutes to construct full disk image of the Sun in the photospheric/chromospheric spectral lines. Figure \ref{fig:spectra} shows an example of spectra of the photospheric spectral lines taken with VSM. ISS is a high spectral resolution (R $\approx$ 300,000) Sun-as-a-star instrument designed to observe in a broad range of wavelengths 350 nm - 1100 nm. The single-lens optical system creates a 400 $\mu$m diameter image of the Sun on the face of a 600 $\mu$m fiber, which is fed to a 2-meter Czerny-Turner double-pass spectrograph in a temperature controlled room. The spectral range is selected using a prism pre-disperser as a pre-filter. Under the current synoptic program, ISS takes observations in nine wavelength bands including Ca {\rm II} 854.21 nm. The spectral sampling of the VSM/ISS in this spectral band is 3.57 pm/1.58 pm (see Figure \ref{fig:spectra2}). 
Additional information about these instruments and the corresponding data reduction can be found elsewhere \citep[e.g.,][]{Bertello.etal2011,Pietarila.etal2013,Pevtsov.Bertello.Marble2014}. Here, we use ISS and VSM observations taken on the same day 
to explore the possibility of reconstructing the disk-integrated (Sun-as-a-star) profile of the Ca {\rm II} 854.21 nm spectral line by combining disk-resolved spectra.

\section{Reconstructing Disk-Integrated Spectra from Disk-Resolved Spectra}

Disk-integrated spectra can be derived as a sum of disk-resolved spectra for each contributing pixel on the solar disk using
\begin{equation}
I(\lambda) = \int^{\pi/2}_{-\pi/2} \int^{\pi/2}_{-\pi/2} I_i(\varphi, CMD, \lambda )d\varphi dCMD,
\end{equation}
\noindent where the disk-resolved spectra $I_i(\varphi, CMD, \lambda)$ at each latitude ($\varphi$) and longitude (central meridian distance, CMD) are weighted by a limb darkening function, $f(\varphi, CMD)$, and their central wavelength ($\lambda_0$) is Doppler-shifted (e.g., solar rotation)

\begin{equation}
I_i(\lambda) = I_{i0}(\lambda_0 + \Delta\lambda(\varphi, CMD)) \cdot f(\varphi, CMD).
\end{equation}

Here, subscript $i$ indicates that the profiles in each pixel may have a different shape depending on the type of feature in that location (e.g., sunspot, plage etc). 

In addition, VSM profiles need to be corrected for a scattered light contribution. We consider ISS profiles to be largely free of scattered light because of the double-pass spectrograph design. The contribution of scattered light to VSM spectra was estimated from the observations of the Venus transit on 6 June 2012. The amount of stray light present was determined {\it a priori} by comparing the apparent intensity of the central region of the transiting Venusian disk to an annular average of quiet Sun pixels outside the Venusian disk.
There was a strong spectral variation to the amount of stray light; about 8\% in the wings, increasing non-linearly to about 13\% in the core. For the analysis presented in this article, we used 10\% contribution from scattered light, independent of wavelength.

To reconstruct disk-integrated spectra from disk-resolved spectra we follow three steps as described below. 

\begin{figure}[h!]
\centering
\hbox{\includegraphics[width=0.4\textwidth]{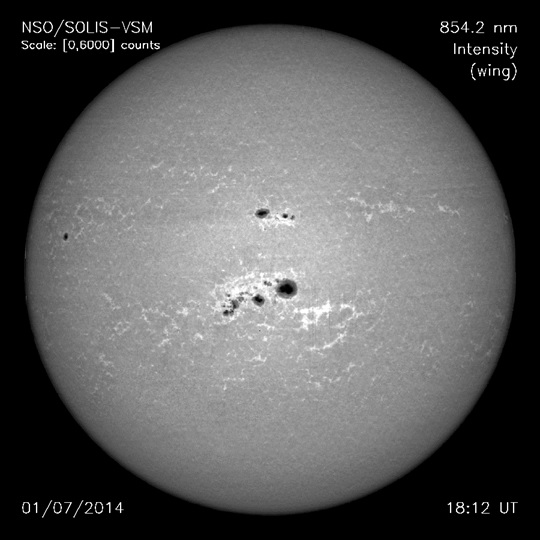}
\includegraphics[width=0.6\textwidth]{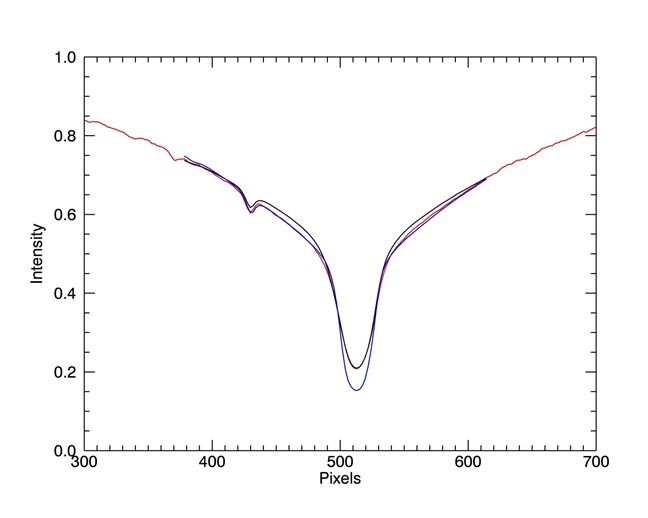}}
\caption{Example of an intensity image derived from the wing of the Ca {\rm II} 854.21 nm  spectral line observed by VSM (left), and a comparison of disk-integrated (ISS, red) and sum of disk-resolved (black) profiles for same day observations. For reference, line profile observed at solar disk center is also shown (blue line).}
\label{fig:step1}
\end{figure}

\subsection{Step 1: Integration of Observed Profiles}

For the first step, we simply integrate all disk-resolved profiles from the VSM. As the intensity in these profiles already reflects the limb darkening, and their wavelength position corresponds to the velocity field on the Sun, these corrections are not needed at this step. Figure \ref{fig:step1} shows that a simple summation of disk-resolved spectra (corrected for the contribution from scattered light) fits the Sun-as-a-star (disk-integrated) spectra quite well. One can note that in the profile computed by summation of disk-resolved profiles (black line) the intensity in the wings is slightly higher than in the sun-as-a-star profile (red line). We believe that this difference cannot be explained by the difference in spectral resolution between VSM and ISS because observed (VSM) profile at disk center does not show such disagreement with the ISS profile. This will be further investigated as we continue this study. 

\subsection{Step 2: Building a Model}

\begin{figure}[h]
\centering
\hbox{\includegraphics[width=0.4\textwidth]{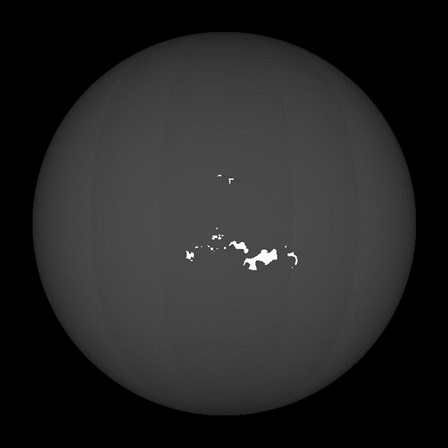}
\includegraphics[width=0.6\textwidth]{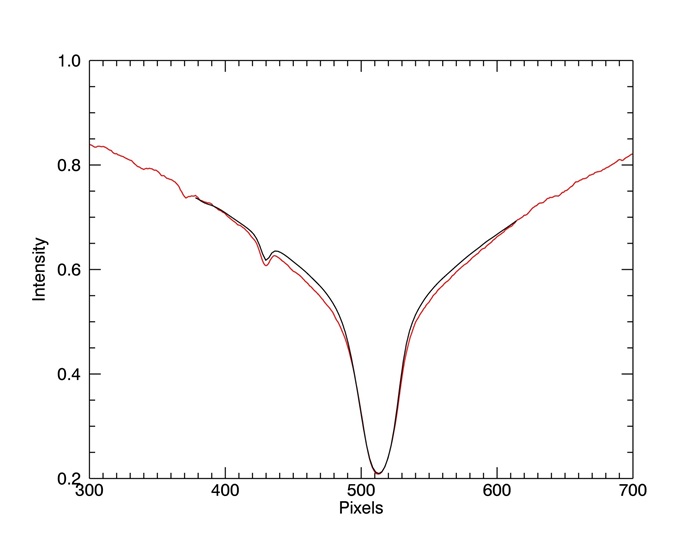}}
\caption{Model intensity image corresponding to VSM observations shown in Figure \ref{fig:step1} left and comparision of disk-integrated (ISS, red) and sum of disk-resolved (black) model profiles.}
\label{fig:step2}
\end{figure}

Next, we replaced all observed (VSM) profiles with ``model'' profiles. To evaluate the number of model profiles that would be sufficient to represent the observed profiles, we analyzed VSM spectra in a small area of the solar disk. The spectra were subjected to a Principal Components Analysis (PCA), and it was determined that two components represent the observed variety of spectral line profiles quite well. Therefore, we created a synthetic data set from the full disk image of the Sun (corresponding to observations shown in Figure \ref{fig:step1}, left) with a single (observed) spectrum corresponding to bright plage areas and single quiet Sun spectrum for all other pixels. The spectra were weighted by a limb darkening function typical of the solar photosphere \citep{Hestroffer.Magnan1998}, and the model spectral lines were shifted in accordance with the solar rotation function. The individual spectra were integrated as in Step 1 to create a single disk-integrated spectrum. The results of this exercise are shown in Figure \ref{fig:step2}. Similar to the case discussed in Section 3.1, the two profiles match each other very well except for the wings.  One can also note that the disk-integrated profile derived by summation of model profiles (Figure \ref{fig:step2}) is very similar to that constructed from the disk-resolved observed profiles (Figure \ref{fig:step1}).

\subsection{Step 3: Interpreting Disk-integrated Profiles in Terms of Plage Coverage}

As the final step, we investigate the possibility of deriving plage coverage (as a fraction of the solar disk) from the Sun-as-a-star data alone.
For this step we used the AutoClass classification code \citep{Cheeseman.Stutz1996} and followed the approach described in \citet{Ulrich.etal2010}. In short:
\begin{enumerate}
\item{} We classify each pixel in a ``training'' set of images by its class. In general, the number of classes can be determined by the code or fixed by the user. For this exercise we limited the number of classes to just two: ``Quiet Sun'' and ``Plage''.
\item{} We establish a model relationship between disk-integrated and disk-resolved parameters. For the particular case of the line core intensity $I_c(t)$ and for each image taken at time $t$, we have:
\begin{equation}
I_c(t) = a + \sum^{J-1}_{j=0} c_jA_j(t),
\end{equation}
\noindent where $J$ is the number of classes ($J$ = 2 in our case), $A_j(t)$ is the expected fractional area of the solar disk covered by class $j$ at the time $t$, and $c_j$ is a coefficient that corresponds to the $I_c(t)$ value the Sun would produce were it covered entirely by a surface of class $j$.
\item{} We apply the model to new data.
\end{enumerate}

\begin{figure}[h!]
\centering
\includegraphics[width=0.7\textwidth]{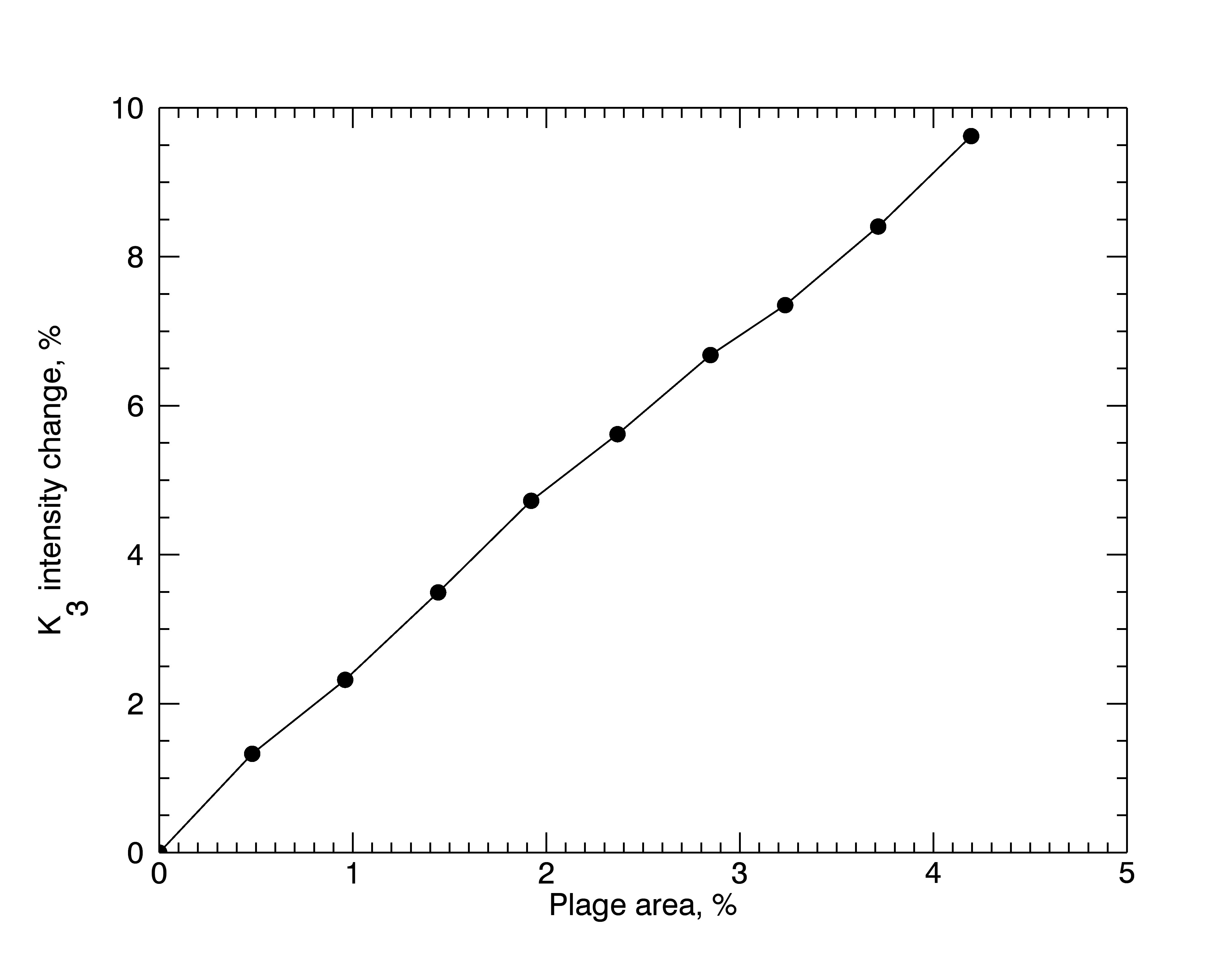}
\caption{Change in core intensity (percentage relative to plage-free solar disk) as a function of area of the solar disk covered by plage, derived from model spectral cubes computed in Step 2.}
\label{fig:autocode}
\end{figure}

As a training set, we employed model data (see Figure \ref{fig:step2}) derived from VSM observations taken on 1 July 2014. We further modified the model data set by varying the area of the solar disk covered by plage. Figure \ref{fig:autocode} shows the change in core intensity of the model Ca {\rm II} 854.21 nm spectral line computed as in Step 2 as a function of the fraction of the solar disk covered by plage.

\section{Summary}

We present results of a preliminary analysis of spectral line profiles of Ca {\rm II} 854.21 nm observed  as Sun-as-a-star and in disk-resolved mode. We show that disk-integrated profiles of this spectral line can be represented reasonably well by a summation of disk-resolved profiles.
Grouping of disk-resolved profiles suggests the prevalence of two classes of profiles (quiet Sun and plage). We also demonstrate that interpreting disk-integrated profiles in terms of the fraction of the solar disk covered by plage can be done via an AutoClass classification approach.

\acknowledgments{
This work utilizes SOLIS data obtained by the NSO Integrated Synoptic Program (NISP), managed by the National Solar Observatory, which is operated by the Association of Universities for Research in Astronomy (AURA), Inc. under a cooperative agreement with the National Science Foundation.
}

\end{document}